**Imaging electrical activity of neurons with metamaterial nanosensors**

Roman V. Beletskiy 2013

**ABSTRACT**

A technology for recording electrical activity of large neuron populations at arbitrary depth in brain tissues with less than cell spatial and millisecond temporal resolutions was the most craving dream of neuroscientists and a long pursued goal of engineers for decades. Even though many imaging techniques have been devised up to date, none of them is capable to deliver either quantitatively valid data nor able to meet contradictory requirements posed for sensors to be safe, non-invasive and reliably working either within cultured cell populations or during chronic implantations in vivo. In my research project, I design and justify a novel nanobiosensors, capable to detect and optically report the electric fields across cellular membrane and investigate properties of that specially engineered plasmonic nanoantennas.
In the following literature survey, I observe the current state of electrophysiology methods and after recalling the basics of fluorescence, discuss benefits and drawbacks of today's voltage sensitive labelling and electric fields imaging methods. This review is wrapped up by a brief outlook of prospective applications of this technology.

**CONTENTS**

**1. Motivating introduction.**

A capability to detect and register biological electric fields is accounted as one of the most important and integral part of the research process in neuroscience domain. Obviously, acutely recorded data of the dynamics of action potentials, slowly varying local field potentials, and electric fields at long and short term potentiation processes at synaptic contacts can shed the light on minute details of neat functional properties and intricate concerted interplay of live neuronal networks.

Two general toolkits are available today to monitor acute neural electrical activity via direct contact with live matter. The most precise for single-cell assay is a patch clamping technique which is supplemented by, perhaps, the most widely used and well established microelectrode array (MEA) recordings, available today in almost all laboratories worldwide.

*Patch clamping* can report on a single cell electrical properties with ultimate precision which is deemed as its most prominent advantage [1]. The resolution of this measurement could be scaled down to gauge the current of specific ions through specific membrane channels [2]. However, many instrumental advantages of patch clamping technique are all levelled off by a number of drawbacks: this method does not allow multiple recordings from bulk tissues (ex vivo or in vivo) thus suitable only to a culture preparations; inevitable cell death within hours after patching membrane due to its irreversible damage; filigree application and individually scored results.

A huge variety of *implantable microelectrode array* architectures [3], devised as up to date can be separated into several general designs: fabricated as long enough and sharply tapered electrodes resided on a flat bed so ready to be inserted from above in tissues (known as Utah electrode array [4] or as flattened shanks with checkered small (of 10 to 30 um in diameter) metal pads on both sides separated from 10 to 100 um each other [5],[6] or uniformly distributed multiple sites on a flat substrate for planar recordings [7], [8], they provide a quite versatile tool as for sensing local field potentials as for acute spikes multi-unit registration at millisecond time scale.

One prominent advantage of that instrument is its capability to sense electrical signals of tens of microvolts at the distance of ~100 μm away the sources [9] thus being the most quantitatively robust extracellular voltage measurement tool to date. However, obvious non-specificity to exciting also juxtaposed neurons in stimulation mode made MEA far from perfect tool to control precise neuronal dynamics. [10]

A necessity for leads and surgical operation required for implantation preclude conventional tethered MEAs from wide use in in vivo setups. A number of *wireless* probes was reported as up to date [11] [12] [13] [14] while their implantable parts are just incarnations of old plain microelectrode arrays, the scale of which neuroengineers are trying hard to push down to allow them to be least traumatically implanted for chronic use. The demand for untethered multi-unit recordings was so high that many neuroengineers around the world put enormous efforts to design microscopic wireless devices to achieve such an ambitious goal. Recent advances and challenges of such miniaturisation efforts reader can find in refs. [15], [16] and [17].

However, careful analysis of theoretical constraints and engineering restrictions imposed on microelectronic devices deemed as an implantable untethered multi-site MEA and tightly packed with a wireless power receiver (generally, a micro-coil inductively coupled to external powering coil), and with a microchip wireless controller (to provide data transmission link to the remote unit), has revealed practical limits in further miniaturisation since its implantable part must be so small that it can not meet a contradictory requirement of miniaturisation against efficiency: the smaller the size of the power receiving coil, the less power could be delivered so the signal amplitude is sacrificed as quadratic function of the single dimension of coil. [18]

To my best knowledge, one of the smallest wireless autonomous MEA neuroprobe was designed by Muller et al. [19] but even this one appeared unable to be introduced into deep tissues because of powering distance shortage in in vivo setups limiting its use only for superficial recordings at cortical areas and unsuitable for voltage volumetric recordings at large. More to the shortcomings of extremely miniaturised MEA, the smaller the electrode's contact area and microwires within implant shall be, the more ambient electromagnetic noise they are readily coupled with. Even though wireless MEAs have a huge advantage over their tethered counterparts, there are other implications barely possible to contend with with a MEA concept, such as a high inflammatory response from tissues due to biological incompatibility of used materials, unavoidable invasiveness of its surgical implantation for in vivo preparations, and its principal incapability to gain discriminated data of activity from large (of 10,000+ neurons) volumes of tissues in vivo with acceptable spiking temporal and spatial resolution.

One possible way to overcome the mentioned above and others untouched limitations of MEA within large-scale network activity recordings framework is to exploit an approach where intra-tissue electric fields are sensed by some agents of perhaps exogenous origin, much smaller than any fabricated microelectronics so that they can be injected with a syringe at any depth or even transported with a blood flow to the target location in a live organism and which are at the same time could be externally interrogated by a wireless means. These requirements are readily met by various types of fluorescent labels widely used as for contrast imaging of biological objects as for detecting intimate events at the cellular and molecular level.

## 2. Literature review

### 2.1 Fluorescence imaging.

Fluorescence imaging is one of the most well established methods to observe the smallest structures of biological species at the highest optically achievable resolution. Basic physics underlying the fluorescence process is depicted on Fig.1:

Generally, conventional *one-photon fluorescence* often suffers as from strong background autofluorescence by organic compounds as from that absorption band of most fluorophores is widely overlapping the emission one [20]. Here comes *two-photon fluorescence* (TPF) handy when two photons if absorbed 'simultaneously' (within ~ 5 fs) by a fluorophore lead to a successive emission of a single photon with the energy slightly less than twice of energy of incident photons, hence, the emitted photon has almost twice higher frequency which makes the fluorescence image easily filtered out of the reflecting background. Once the emission band is to minimally overlap with intrinsic autofluorescence spectra of tissue endogenous fluorophores, TPF turns out to be an invaluable tool with the highest signal-to-background ratio, widely adopted to obtain contrast images of objects labelled with fluorescent dyes or proteins up to the depth of several hundreds of microns [21].

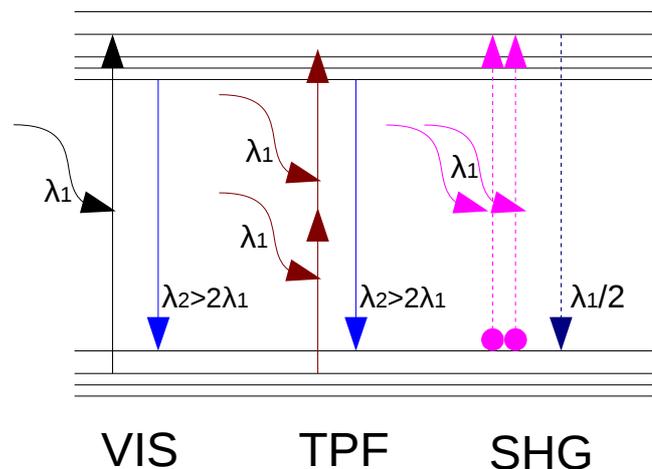

Fig. 1: Schematic representation of fluorescence. VIS – one-photon fluorescence; TPF – two-photon fluorescence; SHG – second harmonic generation.

When two incident photons are being absorbed by a fluorophore within ~5 fs time, this event may also lead to an emission of a single photon at exactly twice shorter wavelength (*second harmonic generation*, SHG) or with thrice (*third harmonics generation*, THG) and so on through to higher energies with decreasing probability. The wealth of mechanisms emergent from the light-matter interactions [22] is not limited to the above scheme , and definitely falls out of the scope of this work. Among many benefits of those high-contrast non-linear imaging techniques like TPF and SHG [23], there are some drawbacks, main of which are high power illumination causing significant photobleaching [24] [25] of fluorophores and tissues photodamage [26] [27]. Of course, the imaging as a methodology is not limited to the above mentioned approaches but alternative methods of visualisation like in [28] [29] of live tissue functional dynamics are out of the scope of this work.

A certain range of wavelengths (700 – 1000 nm) also called as "tissue transparency window", is of special interest for in-depth investigations because water, haemoglobin and oxyhaemoglobin hit altogether their relatively lowest combined absorption [30] (Fig.2) thus making the *near infra-red (NIR)* spectral range most convenient for a tailored instrumentation to look beneath the limits of visible outreach [31] [32]. Even though NIR photons can travel farer than their shorter wavelength 'siblings' due to longer free path in the matter, they finally become elastically scattered which makes direct microscopy impossible beyond the depth of several hundred microns. However, thank to the higher transmittance rate, this spectral range allows for a number of image reconstruction techniques [33] [34] [35] [36] to re-build the original or close o original picture out of backscattered and dispersed non-ballistic photons. Since there is an abundance of optical methods and fluorescent markers tied closely to the optical band used, for the purpose of better specificity and future applications in vivo [37], [38] I will focus in my further work on NIR reporters only.

In general, with an exception of chemiluminescent luciferase based proteins [39] or reversed to light-emitting ex-photosensitive rhodopsins [40] whose ancestry gave rise to the whole optogenetics [41], all fluorescent markers fall into two large categories: *fluorescent dyes* and *fluorescent proteins* whereas the latter could be either of exogenous origin introduced into tissues in pristine form or chemically bound to some delivery vehicle or being often knocked-in into the cell genome so that genetically encoded fluorescent proteins (GEFP) activatable upon certain conditions or being permanently expressed can make the target cell visible under the excitation light.

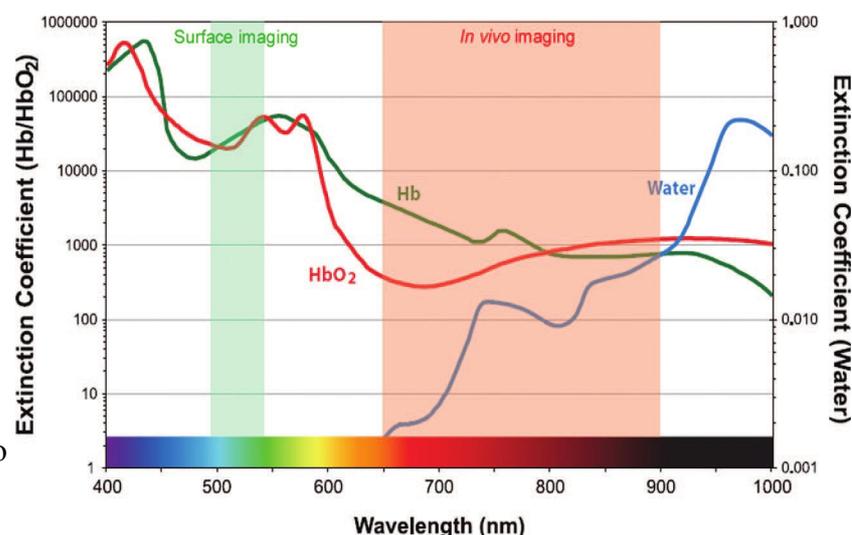

*Fig.2. Tissues transparency window. Cited from Chem.Rev.2010, DOI:10.1021/cr900263j*

A chase to obtain the most robust and reliable dye is yet ongoing [42] since an emergence of latest NIR dyes such as mKate [43] or Si-rhodamine based [44] makes their application even more attractive. The other class of imaging contrast agents, fluorescent proteins possess properties inherited from green fluorescent protein (GFP), firstly observed in jelly fish Aequorea victoria. Many advanced variants are available today [24] [45] however, neither dyes, nor proteins were shown to contend with their primary drawbacks: low quantum yield, relatively slow temporal response time and low photostability that preclude their use as a quantitatively valid tool in capturing precise cellular dynamics. The details of properties and underpinning physics and chemistry could be found elsewhere [46] [47] [48] whilst the best overview of the state-of-the-art in this field of NIR imaging including nanoparticles as fluorescent labels is given in [49] together with an exploratory outlook of the physical and chemical grounds.

Indeed, advent of nanoparticles as optical reporters can be called a breakthrough because unique properties of *quantum dots* (QD) and metal *plasmonic nanostructures* have allowed to overcome many of shortcomings of molecular probes[50]. Nanoparticles of different nature, shapes and compositions are extensively used for contrast imaging of biological species [51], biosensing, and biomedical engineering such as heat-induced destruction of cancer cells [52]

Metal nanoparticles (NP) manifest distinct plasmonic properties that give rise to localised surface plasmon resonances (LSPR)[53]. Due to the fact that resonance wavelength depends on geometrical factors, permittivities of metal and dielectric medium while the scattered light intensity depends on the extinction ratio of NP, the intensity and polarisation of incident light but show no variations upon temperature and no photobleaching, plasmonic NP are extensively used for high contrast imaging [54] [55] and cancer cell targeting [56] [57] [58]while semiconductor quantum dots [59] are mainly targeted as a replacement in situ reporter for molecular fluorescent probes [60] [61]. Other 'exotic' types of nanoscale moieties like fluorescent nanodiamonds [62] or organic polymer nanoparticles [63] [64] etc. fall out of the scope of this report due to their vast variety.

## 2.2 Imaging voltage in neurons with voltage-sensitive dyes and genetically encoded voltage-sensitive fluorescent proteins.

Peterka with co-authors [65] analysed the current status and envisaged challenges one has to meet towards robust and ratiometric optical sensing and reporting of electrodynamics at the neuron membrane. To enable scientists with opportunity to monitor membrane electrical activity, there are voltage-sensitive (also called potentiometric) dyes (VSD) and voltage-sensitive fluorescent proteins (VSFP) that are mainly genetically encoded into genome of specific cell types thus allowing for discriminating monitoring after their activity in vivo.

VSD work pretty similar to other fluorescein-based markers: the chromophore within its molecule undergo significant energy state shift upon changes in electric field the molecule is in so once the field varies, the fluorescence excited under external illumination changes its spectrum. Great overview of VSD use is given in refs. [66] [67] with primary focus on brain imaging.

However, while VSD is most versatile and widely used tool to visualise the electric field in vivo [68] and especially in the brain [69] [70], they have proven to be yet cytotoxic, unstable and susceptible to photobleaching and other adverse effects. In addition, pristine visible and NIR VSDs exhibit non-linear and weak (typically ~10-15% $\Delta F/F$ (light flux) per $\Delta V=100mV$) response, yet slow kinetics, significant capacitive loading to membrane [71]and cytotoxicity.

Another large and yet developed class of neural electrical activity optical indicators are the voltage-sensitive fluorescent proteins. Quite comprehensive and latest outlook of VSFP is given in refs. [72] and [73]. While genetically encoded VSFPs may boast distinct advantage versus VSD due to their capability to label only specific cells in live organism, they appear even dimmer than VSD [74], yet too slow to trace fast neural events like action potential propagation while incurring merely irreversible genetic modifications to the cells of interest that inevitably bans their use in humans.

Although continuous development of novel genetically encoded NIR VSFPs [75] and new variants of NIR VSD [76] [77] promised much better outcomes, both classes of molecular indicators pertain intrinsic drawbacks such as close excitation and emission peaks of quite wide absorption and radiation spectral bells, low quantum yield, photobleaching [78] and volatile optical response function. These disadvantages made inorganic fluorescent labels such as plasmonic nanoparticles and quantum dots the most promising exogenous probes for morphological and functional imaging applications.

## 2.3. Imaging voltage with inorganic nanostructures.

### 2.3.1. Plasmonic nanostructures.
Even simple monomaterial plasmonic metal nanocrystals such as gold nanorods exhibit slight (~5-10 nm) localised surface plasmon resonance (LSPR) shifts upon voltage changes (~0.5V) [79 80 81 82]. Evident from the physics, semiconductor nanocrystals such as ITO with lower concentration of free carriers should see larger LSPR shifts[83] than those in noble metal nanostructures.
Meanwhile, specially designed configurations of nanocomponents of specific geometry and shapes such as nanodisk displaced inside nanoring, give rise to Fano resonances [84] which bring even sharper narrowed down scattering spectra making promise to see to easily discernible LSPR shifts.

### 2.3.2. Semiconductor quantum dots.
There are quite few reports on semiconductor quantum dots used immediately for electric fields imaging. Unsuccessful application of QD as rather straightforward cross-membrane voltage sensors was reported by Invitrogen scientific group in 2008[85]. However, some works where authors used QD-dye complexes exploiting Forster resonance energy transfer (FRET)[86] or big micelles with QDs enclosed[87] give some promise of QD applications towards probing cellular electrical activity.

Summarizing all above, I believe that more complexly organised nanostructures [88 89 90 91 92 93] would allow to achieve the primary objective: devise quantitatively valid voltage-sensitive optical nanoreporters.

## 3. Further development and prospective outlook.

A novel voltage biosensing nanosystem designed primarily for imaging of cross-membrane electric fields in *in vivo* or *ex vivo* conditions, can be further extended to a completely automated optophysiology rig, capable of quantitatively valid cross-membrane potentials measurements via external processing equipment such as dark-field microscope coupled with high throughput spectophotometer and high-speed high resolution CCD camera.
Complemented by the neurostimulation, this novel technology may pave the way to obtain the finest possible neurointerface, capable as to record acute neural electrodynamics from all cells of interest and their synaptic contacts at once as to excite them in precise and controllable manner thus inducing predefined and highly specific activity 3D patterns in live organism.
Other prospective applications of these nanosensing platform could be the real-time screening of local electric potentials in vast variety of artificial and natural materials and tissues.